\documentclass{llncs}
\usepackage{hyperref}

\usepackage{graphicx}

\title{The development and deployment of formal methods in the UK}

\author{Cliff B. Jones\inst{1} and Martyn Thomas\inst{2}}
\institute{School of Computing, Newcastle University, UK\\
\and Emeritus Professor of IT, Gresham College, UK}

\begin{document}
	
\bibliographystyle{alpha}
	
\maketitle

\begin{abstract}
UK researchers have made major contributions to the technical ideas underpinning formal approaches to the specification and development of computer systems. 
Perhaps as a consequence of this, some of the significant attempts to deploy the ideas into practical environments have taken place in the UK. 
The authors of this paper have been involved in formal methods for many years and both had contact with a significant proportion of the whole story.
This paper both lists key ideas and indicates where attempts were made to use the ideas in practice. 
Not all of these deployment stories have been a complete success and an attempt  is made to tease out lessons that influence the probability of long-term impact.
\end{abstract}


\section{Introduction}
\label{S-intro}

The term ``formal methods'' covers the use of mathematically precise notations to specify and to reason about systems.
This paper is mainly concerned with formal methods for software development.
As well as mentioning some of the important scientific insights underlying such methods,
emphasis is placed on the application and deployment of technical ideas and their tool support.

Both of the authors of this paper believe that the use of formal methods is essential for the creation of high-quality software.
This paper is,
however,
not a sales pitch for such methods.
What it sets out to do is to review some actual deployments,
to be frank about drawbacks as well as report successes  
and to identify factors that have either aided or impeded the deployments.

There is no claim that this paper reports on all 
(nor even the most important)
applications of formal methods;
\cite{WoodcockEt09} offers a broader survey.
Given that a selection had to be made,
it felt reasonable to choose applications where the authors had personal knowledge or at least had direct access to colleagues with such knowledge.
There is thus a strong UK emphasis.%
\footnote{Companion papers in this edition of {\em Annals} cover other European countries.}
 
There is no attempt here to claim that formal approaches are a panacea in the sense that they would solve all of the problems in the software industry.
For example,
the question of how to argue that a (formal or informal) specification actually meets the needs of users is a huge challenge --- 
the wise writings of Michael Jackson such as~\cite{Jackson00} are a good starting point on this topic.
Furthermore,
there is the challenge of designing systems that are dependable when used in conjunction with faulty components 
(including hardware and human) ---
in this area~\cite{DIRC-Structure06,romanovsky2013industrial} are useful starting points.


Beginning with the theoretical insights,
Tony Hoare's paper on the axiomatic basis of programming languages~\cite{Hoare69a}
is a foundation stone of a significant portion of the work on formal methods.%
\footnote{An excellent technical survey of 50 years of research on {\em Hoare-Logic} is~\cite{AptOlderog-19}.
An attempt to outline the broader picture is~\cite{Jones03i}.}
In tandem with notations that have been used to provide abstract models of complex systems
(e.g.~VDM~\cite{Jones80a}, 
Z~\cite{Hayes93},
B~\cite{Abrial96} and
Event-B~\cite{Abrial2010})
Hoare's ideas permeate many of the applications discussed in the body of this paper.
Robin Milner's 
LCF~\cite{Gordon79} system provided a model for many of the theorem proving assistants that have been developed.%
\footnote{Historical papers on these topics include~\cite{gordon2000lcf,Paulson2019,Moore2019}
and~\cite{Mac01} looks at what it means to claim that something has been proven.}
UK computer scientists have also made major contributions to concurrency research
(e.g.~Milner's CCS~\cite{Milner80a} and
Hoare's CSP~\cite{Hoare85d}).

Approaches to recording the formal semantics of programming languages is 
another area where UK researchers have made crucial contributions;
although this plays a small part in the rest of the paper,
Section~\ref{S-PLs} outlines this research and identifies some of its applications.

We decided against trying to follow a strict timeline for the paper.
This is partly because the various strands of the story overlap heavily in time
but we also felt that the lessons could be more clearly illustrated by looking at different modes of engagement.
Beyond Section~\ref{S-PLs}, the structure of the paper is that 
Section~\ref{S-cons} considers initiating academic/industry collaboration using consultants,
Section~\ref{S-inside} reports on some projects where the expertise was in-house,
Section~\ref{S-factors} identifies some factors that have an impact on any deployment and
finally, conclusions are summarised in Section~\ref{S-concs}.


\section{Programming languages}
\label{S-PLs}


Although the bulk of this paper is concerned with the specification and development of programs, 
there are two reasons to address research on programming languages:
(i) this was one of the first areas where it was realised that formalism could make a contribution,
and
(ii) UK researchers played a significant part in the development of such research.

The study of languages is sometimes referred to as ``semiotics''.%
\footnote{Charles Sanders Peirce 1839--1914 wrote extensively on philosophy, language and logic 
--- in addition to his collected works,~\cite{peirce1991peirce} reprints an earlier book.}
For programming languages,
the most important facets are syntax (covering content and structure) and semantics (or meaning).
Finding suitable notations to define the syntax of programming languages was achieved both early and with broad consensus.
The ALGOL~60 language was described in~\cite{Naur60} using {\em Backus Normal Form}%
\footnote{John Backus made the initial proposal.}
(often referred to as  {\em Backus Naur Form}).%
\footnote{Peter Naur applied the notation to ALGOL and proposed extensions to the notation.}
Variants of BNF have been developed including {\em Extended BNF}~\cite{Wirth77}
and, also from Niklaus Wirth, a graphical representation referred to as {\em Railroad Diagrams}.
All of these notations perform essentially the same function and there is little doubt as to their usefulness.
One bonus from using such formal descriptions of the syntax of programming languages is that they can be employed to 
generate parsers for the front-ends of language translators.
This does bring some additional concerns about ambiguity and efficiency but, again,
there is broad consensus on how to resolve these more detailed points.

The problems of finding ways to describe the {\em semantics} or meaning of programming languages 
proved to be far more challenging.
A landmark conference was held in Baden bei Wien in 1964.
This first ever IFIP working conference focussed on the subject of 
{\em Formal Language Description Languages} and most of the talks addressed 
proposals for ways to describe formally the semantics of programming languages.
One paper that had considerable influence on the work of the IBM Laboratory in Vienna
was by the American John McCarthy.
In~\cite{McC66},%
\footnote{The proceedings of the conference~\cite{Steel66} took some time to be published but are invaluable to those 
who want to understand the development of ideas because the post-presentation discussions were captured.}
he constructed an {\em abstract interpreter} for {\em Micro-Algol}.
(Essentially,
an interpreter takes a program and a starting state and iteratively computes its final state;
McCarthy used the adjective ``abstract'' to indicate that the metalanguage in which the interpreter was written 
was limited and mathematically tractable.)
McCarthy's paper provided an {\em operational semantics} for a very small language.

The influence of work at the IBM Vienna Lab on UK research becomes important below. 
The Vienna work in the 1960s saw the ideas of operational semantics extended and applied to the huge PL/I programming language.
Their techniques became known as the {\em Vienna Definition Language (VDL)} ---
see~\cite{LucasWalk69}.

The task of formally describing the evolving PL/I language was separate from the language design team.
Of course, there was strong interaction and communication.
But, whenever the formalisation detected problems in the inherently ambiguous
(and sometimes contradictory)
natural language description,
the formalists had to communicate the problem
(sometimes by writing an indicative program).
The response was then an amendment to the text which again might not be crystal clear.
The {\bf lesson} here is that separation of designers from formalists is far less productive than working as an interdisciplinary team.

Turning to key UK speakers at the 1964 working conference,
both Christopher Strachey 
and Peter Landin spoke about a more abstract approach than operational semantics:
rather than define an abstract interpreter,
they were proposing that the constructs of a programming language  should be translated into mathematical functions.
Such functions were written in a notation known as the Lambda calculus.
The development of what later became known as {\em denotational semantics},
is a fascinating story ---
suffice it here to say that researchers at Oxford University made the seminal contributions.%
\footnote{An event was held in Oxford in November 2016 to mark the centenary of Strachey's birth and
videos of the talks are available at:\\
 http://www.cs.ox.ac.uk/strachey100\\
An excellent biographical note on Strachey is~\cite{Cam85}.} 

At the beginning of the 1970s, 
researchers at the IBM Lab in Vienna also made the step from operational to denotational semantics.
The PL/I language was again a catalyst:
the Lab had been invited to build the PL/I compiler for a radically new machine architecture.
Unfortunately these machines were never built, 
but the aspect of VDM 
(the {\em Vienna Development Method}) 
that related to describing language semantics was rescued by writing~\cite{BjornerJones78,BjornerJones82}.
A more detailed account of this work can be found in~\cite{JonesAstarte-16TR,AstartePhD}
and the step from VDL to VDM is discussed in~\cite{Jones01d}.
Here, again, there was a damaging wall between language designers%
\footnote{The source language of the compiler was to have been that of the evolving ISO standard
and the ISO PL/I standardisation committee were still evolving the language.}
and the formalisers;
moreover, the same mistake was repeated on a formal description of the machine architecture itself.
In both cases the separation proved wasteful and far less effective than if there had been a more tightly knit structure.

Returning to the Baden-bei-Wien conference,
Tony Hoare expressed unease about all of the proposed techniques because he saw the need to leave some aspects of a language undefined.
Obvious cases include features of a programming language that relate to implementations on specific machines.
But Hoare's interjection was prescient because concurrent programs can legitimately yield different results
depending on the rate of progress of their separate threads.
This observation led Hoare to develop an {\em axiomatic approach}~\cite{Hoare69a} 
which is key to reasoning about programs satisfying formal specifications.

The challenge of describing 
concurrent programming languages with their inherent non-determinism
was overcome by Gordon Plotkin's 
{\em Structural Operational Semantics (SOS)}~\cite{Plotkin81}.%
\footnote{These lecture notes from 1981 were widely circulated and, thankfully,
reprinted as~\cite{Plotkin03a};
they are accompanied by a useful commentary~\cite{Plotkin03b}.}
 
 The history of formal semantic descriptions is covered at length in Troy Astarte's doctoral thesis~\cite{AstartePhD} 
 and more technical details are given in~\cite{JonesAstarte-16TR,JonesAstarte-SC-proc}.
(The specific formalisation of the Spark-Ada language is discussed below in Section~\ref{S-inside}.)

 

\section{Consultant led deployments}
\label{S-cons}


One obvious mode of transferring ideas from academic originators into industrial practice is 
for the originators to act as consultants to practitioners.
This offers a way of overcoming the inevitable lack of knowledge of novel ideas in the receiving organisation 
but it runs several risks as outlined in the lessons spelled out at the end of this section.
Specifically for formal methods ideas, 
this approach runs the risks associated with separation of practitioners from formalists mentioned in Section~\ref{S-PLs}.
Probably the best-known of these deployments was the work at IBM on their CICS system
(description below)
but there are also some useful lessons from a less well known activity in STL
that is covered in Section~\ref{S-inside}.

The first issue is,
of course,
how to initiate the contact.
%
%
Before coming to CICS itself,
some background activities are worth outlining.
The aspects of VDM associated with the semantic description of programming languages
are mentioned in Section~\ref{S-PLs} but VDM is more widely known as a development method 
for (general) programs.
Early work on the program specification and development aspects of VDM was actually undertaken inside IBM at the UK Laboratory in Hursley~\cite{Jones71c,Jones73a};
the first book on this aspect of VDM was~\cite{Jones80a}.

Starting in the 1970s,
there was a programme of 
{\em European Laboratories Integrated Professional Training Program} 
(ELIPT)
technical courses.
A course based on the 1980 VDM book was offered and taught by Derek Andrews and Cliff Jones.
Management at the IBM Laboratory in Germany at B{\"o}bligen made the decision to enrol most of their active programmers on 
this course and employees attended in more-or-less coherent project groups.
A typical course would begin at a hotel in the {\em Schwarzwald}
with two weeks of lectures and writing exercises 
followed by a one-week intensive workshop that initiated a formal description of a 
(simplified version of) the product of interest to that group of engineers.%
\footnote{Jones was involved in teaching of the first eight ASD courses;
the majority took place in Germany, a couple in the UK and one in Italy.}
Managers claimed that they were too busy to commit to this length of course and 
a shorter
{\em Management Awareness} course had to be tailored to their needs.
There were several significant success stories:
one that is described in an external publication is~\cite{beichter1983slan}.

In contrast to this organised enrolment,
the IBM development laboratory in the UK at Hursley
simply let individuals from any project enrol on the ELIPT course
in a fairly random way.
This meant that when needed
(see below)
there was no critical mass of engineers who were all up to speed on VDM.
The {\bf lesson} here is that education needs to establish a cohort of people in the receiving organisation.

To emphasise this lesson it is worth comparing with Harlan Mills' success in IBM's 
{\em Federal Systems Division} (FSD) during the early 1970s.%
\footnote{This is reported from Jones' memories of several personal discussions with Mills.}
Mills persuaded the director of FSD to attend the first course on his formal approach which ensured that no intermediate managers 
could claim they were too busy to take part.
There was also a notion of ``passing'' the course.
Effectively,
most development engineers in FSD attended.

The story of using formal methods on CICS began in 1980 and ran until 1993.
During 1979--81, 
Jones was doing his (belated) Doctorate under supervision of Tony Hoare at Oxford University.
Jean-Raymond Abrial arrived at the {\em Programming Research Group} at the same time as Jones.
Hoare arranged that the two shared an office and many interesting discussions were  conducted with Jones writing specifications in the VDM notation from~\cite{Jones80a} 
and Abrial experimenting with what was to become the Z notation (see below).


IBM's {\em Customer Information Control System} (CICS)%
\footnote{See https://en.wikipedia.org/wiki/CICS.}
is an on-line transaction processing system used by major financial and retail organisations.
CICS had evolved from a customer's 1968 program to be a full-blown IBM product.
In 1988 it consisted of well-over half a million lines of code written in at least two languages.
This was the period of software ``unbundling'' and
it was believed that CICS was at one time one of IBM's most profitable program products.

Because of his contacts with IBM Hursley%
\footnote{Jones had worked in Hursley 1965--68, 1970--73;
the gaps filled by assignments to the IBM Lab Vienna and IBM's {\em European Systems Research Institute} in Belgium.}
and his on-going courses,
Jones was asked about ways to help the CICS team adopt formal specification.
Starting in 1980 there were informal discussions that led,
in early 1981, to the suggestion of a contract with a university
(i.e.~not with single consultants).
Formal meetings between Hursley managers and Prof Hoare in the middle of 1981
resulted in a contract between IBM Hursley and Oxford University being in place by year end.
%
Ib S{\o}rensen and Tim Clement
were to work on the project from the Oxford end;
Hursley people included Pete Collins, John Wordsworth and Peter Lupton;
Hoare had asked that Jones worked as a consultant on the project%
\footnote{Having submitted his thesis in June 1981
Jones moved to a chair in Manchester University starting August.}
and Rod Burstall of Edinburgh University was involved in the same role.

As mentioned above,
Abrial was developing the ideas that coalesced into the Z specification notation
and it is clear that the challenges presented in describing CICS had an influence of this evolution.
Key discussion partners also included Bernard Sufrin and
Carroll Morgan.
Ian Hayes joined the project in January 1983, 
which was perhaps the key time for the development of Z's so-called ``schema calculus''.
Hayes also went on to edit the first book on Z~\cite{Hayes87a}.%
\footnote{The frustration at not having a stable reference document for Z in 1981 
led to the construction of a spoof document with Sufrin's name shown as author
but actually comprising a pastiche of other papers put together by researchers at PRG .}
Other Z books from around this time include~\cite{Spivey88,Spivey92,Wordsworth92,mcmorran1993} and Mike Spivey also programmed the first tool that type-checked Z specifications.
%
A useful intermediate report on ``CICS Experience with Z'' is the (unrestricted) Technical Report~\cite{TR12.260}. %
The summary includes the following positive assessment:

\begin{quotation}
\noindent
``From the industry point of view, this work has demonstrated that:
\begin{itemize}
\item provided that there is adequate education and support,
a mathematical notation such as Z can be used for software development in an industrial environment;
\item the use of formal methods changes the development process and brings greater precision to earlier stages;
\item communication between development groups can be improved.''
\end{itemize}
\end{quotation}
Jim Woodcock worked on the Oxford end of the collaboration from 1985--93.
In particular he sorted out the logic underlying Z~\cite{woodcock1992w};
he also co-authored a book on Z~\cite{woodcock1996using}.
Steve King joined the project in January 1986 and stayed to the conclusion of the contract in 1993
(he spent a year working in Hursley (1990/91).
The second edition of Hayes' {\em Case Studies} book~\cite[Part-IV]{Hayes92a} contains five chapters
by Hayes and King on details of the description of parts of CICS.

%
%
%
%
%

\begin{figure}
\begin{center}
 \includegraphics[width=80mm]{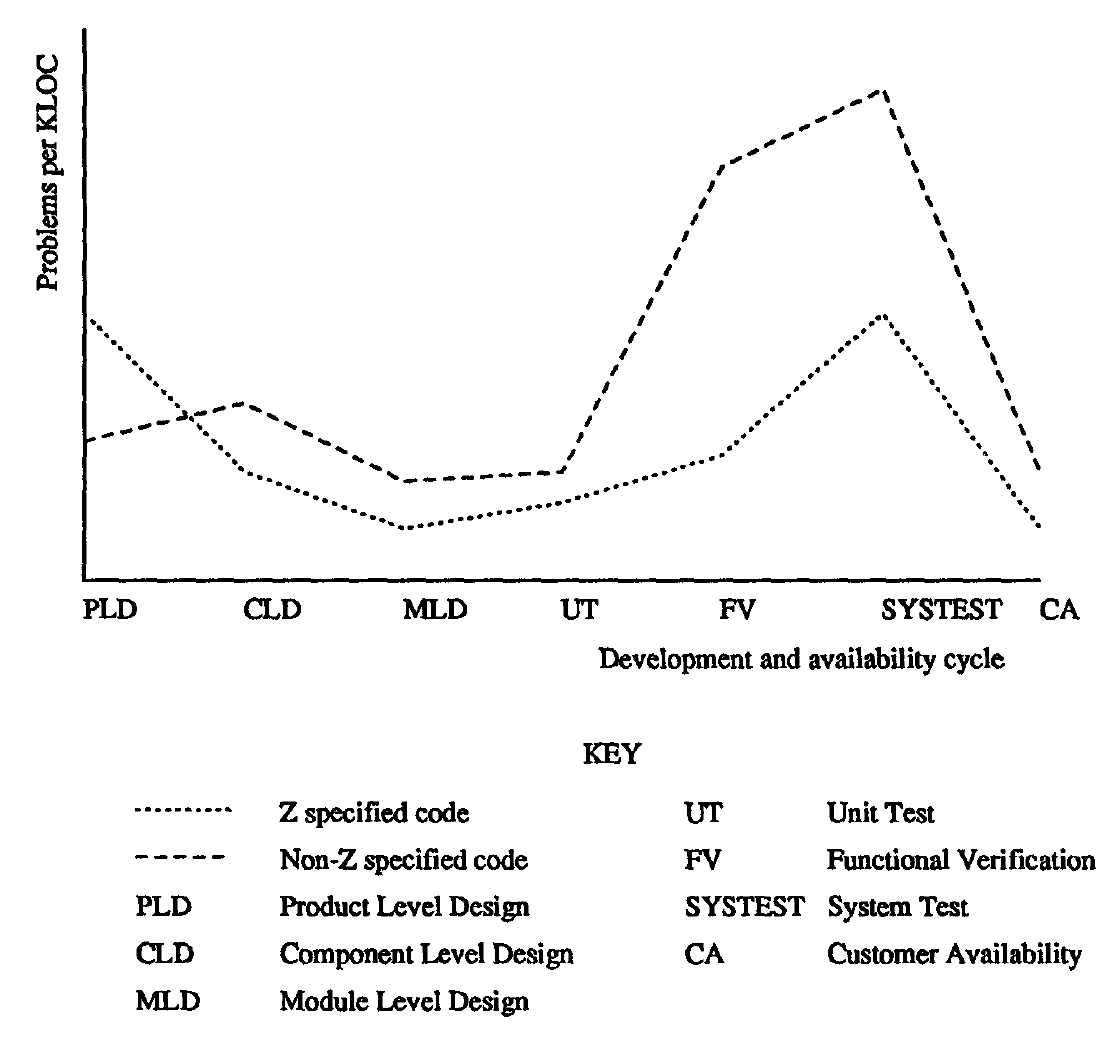}
\end{center}
\caption{Figure reproduced from~\cite{HoustonKing} indicating errors found at different development stages.
[Original caption] Comparison of problems found with two development methods for CICS/ESA V3.1}
\label{F-CICS}
\end{figure}

A joint paper (Steve King -- Oxford University's {\em Programming Research Group} (PRG) and  Iain Houston -- IBM)~\cite{HoustonKing}
provides a strongly positive assessment of the exercise in formalisation.
In particular,
they address an oft-claimed benefit of formal methods  which is that more problems are detected early resulting in significantly reduced problems
--and thus less costs-- 
overall.
The figure reproduced here in Fig.~\ref{F-CICS} provides strong evidence for the validity of this claim from a substantial industrial exercise.
%
%
The collaboration was recognised with a Queen's Award for Technological Achievement in 1992 to the Oxford PRG group and IBM Hursley. 

Despite all of these positive indicators,
an informal reunion of Hayes, Hoare, Jones and S{\o}rensen
(kindly arranged by Jonathan Lawrence) 
in September 2011 found little trace of the continued use of Z in IBM Hursley.%
\footnote{Jonathan Lawrence drew Jones' attention to~\cite{hoareJ1996applying}.}
This prompts an evaluation of the {\bf lessons} from what is widely viewed as one of the success stories of formal methods deployment.

\begin{itemize}

\item PRG staff who were funded by the collaborative contract had to spend significant amounts of time 
working on details of the CICS code;
 
\item not only were general courses on Z designed and given to IBM engineers, but additional ``readers' courses'' were needed;

\item the presence of ``internal supporters'' was crucial at the time of strongest interaction;

\item there was a period when IBM engineers and management were pushing hard for a recognised standard for the Z notation;%
\footnote{An international standard for Z was approved in 2002.}

\item (and linked to the previous point)
there was a perceived need for tools that helped creation, maintenance and analysis of Z documents; %

\item management support is crucial 
(see above on Mills and IBM FSD)
--- the attitude of IBM Hursley management ranged from supportive to antagonistic;%
\footnote{Jones' notes made at the time record one fairly senior manager responding to ``it's just mathematical notation''
with ``I hate mathematics''.}

\item King and Jones ascribe the drift away from the use of Z in Hursley down to people:
there were probably not quite enough internal supporters, 
some moved on to other roles;
less positive managers became responsible for decisions which affected the selection of methods to be used in later releases.

\end{itemize}




Turning to another deployment,
the software for the Sizewell-B nuclear reactor reinforces one of the key lessons.
It was an extreme example of separation of development from verification and the responsible teams.
Sizewell-B was the first nuclear reactor in the UK to have a programmable primary protection system (PPS). 
The software had been developed by the US company Westinghouse without a formal specification; 
there were about 100,000 lines of unique executable code.%
\footnote{It was written in PL/M-86 with some ASM86 and small amounts of PL/M-51 and ASM51.}
Originally, there were two specification documents, a high level Software Design Requirements (SDR) and a more detailed Software Design Specification (SDS). 

The regulator (Nuclear Installations Inspectorate) decided that it was necessary for the PPS software to be formally verified.
The chosen route to (partial) formal verification was to manually translate (in the UK)  the SDR and SDS into a mathematical specification and to write a translator from PL/M-86 to the Intermediate Language (IL) used as input to the MALPAS static analysis tool. This work started in January 1989 and was completed in 1993. The MALPAS analysis project team grew to more than 80 people and the project cost GBP 7 million. A review by Nuclear Electric~\cite{Fenney-95}
 concluded that ``The costs of the MALPAS review are high largely because the specification documents had to be manually translated into a formal notation before they could be used. This leads to the conclusion that the review processes need to be considered during the design phase of the project wherever possible''.%
\footnote{Ward~\cite{Ward-93} goes into detail on the MALPAS use of {\em Compliance Analysis} 
and explains that there was no choice but to develop pre and post condition specifications of components bottom up from the code.}

This clearly reinforces the lesson that a separation of developers and formalists is damaging in general and that leaving verification to the late stages of development is lengthy, difficult and expensive.



An extremely important vector of formal methods research and deployment centres around Jean-Raymond Abrial;
this story links to the UK (and involvement of the current authors) but is actually more international.
After laying the groundwork of what became the Z notation,
Abrial returned to France and acted as an independent consultant.
He not only made a fundamental shift to create the B-method~\cite{Abrial96}%
\footnote{This book contains a generous acknowledgement to the influence of VDM on B.}
but he also developed a ``B-tool'' under contract to BP.
Subsequently,
Abrial developed a completely new ``Atelier-B'' which was subsequently supported by the French formal methods companies ClearSy and Systerel.
The notation and tool were used in the important development of the software of Metro line 14 for RATP~\cite{B4RATP}


Abrial is a fascinating person who deserves a full biographical article.
He is self-critical in the best possible way and has undertaken several complete rethinks of his ideas.
His next step was strongly influenced by books on {\em Refinement Calculus} by
Carroll Morgan~\cite{Morgan94} on the one hand and
Ralph Back and Joakim von Wright~\cite{BackWright98} on the other.
Abrial's {\em Event-B} is described in~\cite{Abrial10}.
Tool support was designed and built in an EU-funded project known as {\em Rodin}
which was led by Newcastle University.
The central tool activity was undertaken by Abrial, Laurent Voison and Steffan Hallerstede at ETH Zurich 
in the chair of David Basin.
 A subsequent EU-funded project ({\em Deploy})%
 \footnote{Other partners included SAP and Bosch ---
the Advisory Group was chaired by Thomas.} 
was again led by Newcastle and this time involved four industrial  companies.
As its name suggests,
the emphasis here was on deployment of {\em Event-B} and the {\em Rodin Tools}.
The most accessible description is~\cite{romanovsky2013industrial}.

%
%

Although neither of the current authors were directly involved,
it would be remiss in this section not to mention the work on hardware verification led by Mike Gordon of Cambridge ---
Larry Paulson wrote the Royal Society report~\cite{Gordon-obit}.%
\footnote{An invaluable resource on theorem proving efforts is~\cite{Mac01}.}
Furthermore,
the Oxford work on exploiting CSP by providing the FDR tool is to be described in a paper being written by Bill Roscoe and Steve Brookes.


\section{Expertise within the deployment organisation}
\label{S-inside}
 
This section considers the use of formal methods in some organisations that made them a more integral part of their development process.

Most if not all organisations that adopt formal methods will have progressed in stages of increasing rigour through informal and then structured methods, informed by their interactions with more advanced organisations and academics. One such organisation is the software engineering company Praxis (later Altran UK) founded by one of the authors of this paper (Thomas) with his colleague David Bean.

Thomas had worked on the design of a computer based PABX at Standard Telephones and Cables (STC) in north London in 1975/6 where he was involved in introducing and teaching functional decomposition using the diagrammatic Structured Analysis Design Technique (SADT)~\cite{SADT}. 
Bean had worked in a leading UK software house, Logica, and was familiar with Jackson Structured Programming (JSP)~\cite{Jackson75}. 
They came together to set up the South West Universities Regional Computer Centre (SWURCC) where Thomas recruited a small team to develop an Algol68 compiler that was required by SWURCC's user community to run on the ICL 2980 mainframe computer.

The Algol68 compiler used a front end, Algol68RS~\cite{WoodwardBond-83}
developed at the Royal Signals and Radar Establishment (RSRE) in Malvern, UK by a team in the Mathematics Division that included Susan Bond, Ian Currie, John Morison and Philip Woodward.  The Mathematics Division at RSRE ran the establishment computing service, having developed the UK's first solid-state computer and written its operating system and compilers. 
RSRE also developed cryptographic systems, 
formal static analysis tools to help identify trojan code and 
provably secure hardware systems (the VIPER processor).
The VIPER activity 
--for example~\cite{Cohn88}--
is an interesting story in its own right:%
\footnote{See~\cite[Chap~7]{Mac01} on VIPER;
Donald Mackenzie's whole book contains a thorough and deep analysis of the concept and limitations of formal proofs about software.}
it used a formal hardware design and development system comprising the Electronic Logic Language (ELLA), developed by Morison and later released under a public license~\cite{Ella-93}
supported by tools for design transformation, symbolic execution and formal verification. 
The SWURCC team impressed RSRE sufficiently that in addition to Algol68, they were commissioned to support and market ELLA.

SWURCC had built a reputation for software quality that led to them being invited to join a consortium ({\em Augusta}) funded by the UK Department of Trade and Industry to investigate the use of the new Ada programming language. Augusta was led by Tim Denvir of STC's telecommunications laboratory STL, who recalls that 

\begin{quotation}
\noindent
``Our report, delivered in September 1981, took a few example problems, expressed a design following several different methods, and developed implementations from each in Ada. We also did a literature study of many more design methods and of developers. Among the mostly structured methods (such as JSD) we used and/or considered CCS and VDM''~\cite{Denvir-17}.
\end{quotation}

The SWURCC team became the kernel of Praxis, set up by Thomas and Bean in 1983, transferring the ELLA and Algol68 compiler support and other projects from SWURCC including a Unix re-implementation for ICL and a FORTRAN compiler for a military version of ICL's Distributed Array Processor.

Denvir 
(the rest of this paragraph draws heavily on Denvir, op.~cit.) explains that STL had significant history in the use of formal methods: by the mid 1970s they were already using Dijkstra's pre and post conditions to prove small programs correct. In January 1979, Denvir attended the Winter School on Abstract Software Specifications in Denmark with an STL colleague, Bernie Cohen, where they met Dines Bj{\o}rner, Cliff Jones, Steve Zilles, Joe Stoy, Peter Lucas, Peter Lauer, Barbara Liskov, Gordon Plotkin, Rod Burstall, David Park, O-J Dahl, Peter Mosses and others. Denvir and Cohen were particularly impressed with lectures on VDM and in 1980 persuaded STC management to hire the services of Cliff Jones as a consultant to apply VDM to telecommunications projects. With sponsorship from Jack Shields, STC's Group Technical Director, STC developed their own courses on VDM and discrete mathematics. These were originally taught by Cliff Jones, then delivery was taken over by Tim Denvir, Roger Shaw and Mel Jackson. Over 2-3 years they trained over 200 engineers from various group companies~\cite{JacksonDenvirShaw-85}.
They experimented with Z~\cite{Hayes87a} in one project using consultancy from Bernard Sufrin and Carroll Morgan from Oxford's PRG. With the support of the LFCS at Edinburgh University, hiring Mike Shields, Denvir's group developed an interest in the use of formal techniques for concurrent systems, and this culminated in a workshop held in Cambridge in September 1983~\cite{Denvir85}
and to the use of Robin Milner's Calculus of Communicating Sequential Processes (CCS)~\cite{Milner80a} in revising the standard for the telecommunications industry standard design language SDL.%
\footnote{https://www.itu.int/rec/T-REC-Z.100/en.}

When the management changed at STC, Tim Denvir and Mel Jackson joined Praxis, bringing with them a European Commission contract to specify the  {\em Portable Common Tools  Environment} (PCTE) in an extended version of VDM (see~\cite{middelburg1989vvsl,Middelburg90} for VVSL).
Praxis was a fertile environment for the introduction of formal methods. They abandoned SADT and adopted VDM~\cite{Jones80a} 
and later Z~\cite{Spivey89}
as specification and design methods, particularly for safety or security critical projects such as the protection system for a radiotherapy machine, the certification authority to support the  MONDEX smart card~\cite{HallChapman-02}, 
a system (SHOLIS) to support the landing of helicopters on naval ships~\cite{king2000proof},
and a system (CDIS)~\cite{Hall96}
to support the UK National Air Traffic Services (NATS) controllers handling aircraft for the five London airports.

Static data flow and program analysis tools had been developed at RSRE based on research by Bob Philips of RSRE. The work was declassified in the 1970s so that it could be used on civilian safety critical projects, resulting in two commercially available products, the Malvern Program Analysis Suite (MALPAS)  and the Southampton Program Analysis Development Environment (SPADE, which became SPARK, see below) developed by a team led by Bernard Carr{\'e} who were doing research in graph theory at Southampton University. Both MALPAS and SPARK have been used in the development and verification of a range of safety critical and security critical systems.

SPADE was originally developed to analyse and verify programs written in a small subset of Pascal but Ada was then chosen as the foundation for future work. An Ada subset (the SPADE Ada Kernel or SPARK) was defined informally by Bernard Carr{\'e} and Trevor Jennings in 1987 and more formally defined using a variant of Z in 
{\em The Formal Semantics of SPARK}~\cite{MarshONeil1994}. 
Bernard Carr{\'e} set up a company Program Validation Limited (PVL) and later sold it to Praxis, with the PVL staff becoming Praxis employees. SPARK has kept up with revisions of Ada as each has been standardised and some industrial applications have been described by Chapman
and Schanda~\cite{ChapmanSchanda-14}. 
A survey of Praxis/Altran projects has been published by White, Matthews and Chapman~\cite{WhiteEt-17}. 
The Critical Systems Division of Praxis was acquired by Altran UK in 1997.

It is important to stress that the Praxis' use of VDM, Z, CCS, and CSP was unusual if not unique. Most groups using formal methods used them to specify small, critical components; in Praxis these methods were used to specify system-level behaviour. This had a profound effect on the benefits and drawbacks that we found.
These are some {\bf lessons} from Praxis' experience with formal methods:

\begin{itemize}

\item Praxis developed its own courses to teach its software engineers to use VDM and Z. The courses lasted four days, preceded by a single day teaching discrete mathematics. This was found to be enough for computer science graduates to be able to read and start understanding a formal specification, though the ability to write good Z developed over a period of working on a project with access to experienced Z practitioners. 
\item formal methods had to be fitted into an overall development process and combined with other techniques, for example prototyping for user interfaces,  and Data Flow Diagrams for process design. Further, no single formal method covers every aspect so Praxis had to use different techniques for functionality and concurrency and find ways of integrating these. They also had to write specifications at the system level and at the design level and show that the design was consistent with the system specification. Using high-level, set-theory based languages meant that there were almost no tools available and this limited their ability to generate executable prototypes, or carry out proofs, model checking or automatic code generation.
\item The benefits of this approach were a systematic, rigorous and traceable development leading to systems with few defects in service~\cite{amey2002correctness,Spark-Hercules,hall2002correctness,king2000proof};
as Praxis improved their understanding of how to do this, they steadily drove down defect levels~\cite{HallChapman-02,Hall-05}.

\end{itemize}

Commercially, using formal methods had drawbacks as well as benefits:
 \begin{itemize}
 \item Customers were nervous about formal methods; the US {\em National Security Agency} asked Praxis to demonstrate the practicality of formal methods by developing an experimental system to control a secure enclave. The Tokeneer project was a success~\cite{Barnes-06b}
in that the NSA were unable to find any faults in the software, Praxis were able to train two NSA interns to extend the system, and the NSA said that the productivity of the Praxis team was the highest they had ever experienced but somehow this did not lead to sales of the SPARK toolset or to further NSA projects. The NSA later agreed that all of the Tokeneer specification, design and development could be publicly released including all the tools and proofs,%
\footnote{https://www.adacore.com/tokeneer.}
so that anyone could download and study the project, experiment with the proof technology and see whether other tools might reveal defects that had not been found. The various follow-on projects and experiments have been summarised in a book chapter by Jim Woodcock~\cite{WoodcockEt-10}. 
 \item on the CDIS project for NATS, Praxis developed a formal VDM specification during the bid, to determine exactly what the customer meant by their requirements. This led to over 100 requests to NATS for clarifications but it enabled Praxis to feel comfortable bidding for a large project whose cost exceeded Praxis' annual turnover%
\footnote{Thomas remarked to Jones over a dinner in Brussels that he had ``bet the company on VDM''.} 
and to contract to repair at no charge any major faults that developed over the following five years. 
NATS later said that CDIS had been by far the easiest system to integrate that they had experienced.
Furthermore its stability was so good that,
after a few years,
NATS had to ask Praxis to retrain their staff in how to restart CDIS because it had failed so rarely that they were unsure how to do it.
 \item The functional specification for CDIS 
 --and subsystem specifications for the two component parts (server and workstation)-- 
were expressed in VVSL.
The user interface used state transition diagrams and for modelling concurrency used CSP and for the LAN design used Milner's CCS. Praxis' technical architect, Anthony Hall, has described and explained the choice of methods~\cite{Hall96}.
  With NATS' agreement, Praxis made the CDIS code and project records available for analysis by two academic researchers whose conclusions~\cite{PfleegerEt-97} 
were that formal methods can contribute to achieving very reliable code (but with many reservations, for which see the cited paper).

\item On CDIS it was extremely useful to have a formal specification and traceable design documents available throughout the project for dealing with life cycle issues such as requirements changes (there were hundreds) and fault analysis. For a given requirements change request it was possible to trace through from the formal specification to identify areas of the implementation affected by it and manage change in a cost-effective and efficient manner. With fault analysis, it was possible to trace back to the formal specification and identify whether the fault was due to an implementation or a requirements error. The use of a formal specification reduced costs and increased efficiency throughout the implementation life cycle.

\item Formal methods projects of this sort need considerable work before any benefits are seen, which made it difficult to convince potential customers to invest in such projects.
\item The lack of support tools can be a major problem for the use of formal methods that only address the specification stage. Some form of prototyping is very important so that clients can validate the specification and request changes while it is still relatively inexpensive to change it.
\item System specifications need to be understood by domain experts as well as by computer scientists. Praxis rewrote the informal specifications using and including the formal specification so that it could be understood by clients. Preparing a formal specification facilitates writing a better structured, clearer and shorter specification in natural language.
\item Praxis offered some free warranties for projects that had a formal specification agreed with the client, as this made it possible to distinguish between errors and specification changes. This was less commercially risky than it might seem, because the warranty would be voided if the client changed the software themselves or asked another company to implement new features.
 \item NATS subsequently contracted Altran UK to develop a new ATC system iFACTS%
\footnote{https://nats.aero/blog/2013/07/how-technology-is-transforming-air-traffic-management/.}  
that was successfully delivered using Z and SPARK. Thomas was present at one NATS meeting before this contract was awarded where the objection was raised that if NATS presented their regulator (The UK {\em Civil Aviation Authority}) with SPARK code and formal proofs as part of the evidence for the safety of iFACTS, there was a risk that the CAA would always require such strong evidence in future!
 \end{itemize}



\section{Some influential factors}
\label{S-factors}

Apart from the lessons for those wishing to deploy formal methods,
there are exogenous factors that have affected the growth and adoption of formal methods.
This section covers some of these influences.

\subsection{Research funding}

In the UK,
funding from its research councils has been supportive to both the underlying research and deployment of formal methods.
In fact, it can be argued that such funding was pivotal in the 1970s: %
viewed from Manchester University (where Jones was at that time) it facilitated the creation of a world class formal methods activity.
It is also worth noting that BP's ``Future ventures'' programme provided funding for activities in 
Edinburgh University's 
{\em Laboratory for Foundations of Computer Science} (and for Edsger Dijkstra).

The body principally responsible for funding computer science research in U.K. Universities was then known as SRC (Science Research Council). 
The Computer Science committee, recognising the importance of distributed computing as a research area, appointed a panel in June 1976 under the chairmanship of Prof. I. Barron, to consider what action was necessary to encourage, coordinate or direct research in Distributed Computing. The  Distributed Computing Systems programme started in the academic year 1977-78. DCS was the first attempt by SRC to establish a long term, extensive, coordinated programme of research in Information Technology. The Technical Co-ordinators of DCS were Bob Hopgood (1977-79), Rob Witty (1979-1981) and David Duce (1981-1984).
The primary scientific objectives of the programme were to seek an understanding of the principles of Distributed Computing Systems and to establish the engineering techniques necessary to implement such systems efficiently. These broad objectives reflect the relative immaturity of the subject when the programme was founded. In particular the programme sought to establish an understanding of parallelism in information processing systems and to devise ways to take advantage of this.
When the DCS programme was first established, the research covered five major topic areas, representing a progression from fundamental theory to novel applications. The areas were:
\begin{itemize}    
\item Theory and Languages: An adequate theoretical basis for Distributed Computing Systems. 
\item Resource Management: Distribution of control, allocation, scheduling and organization. 
\item (Machine) Architecture. 
\item Operational Attributes: Particularly reliability and performance. 
\item Design, Implementation and Application: Hardware and software techniques for development and implementation.  
\end{itemize}

A  major theme in DCS was concerned with theories of parallel computation and with the development of notations and techniques for specifying and verifying such systems.%
\footnote{See \url{http://www.chilton-computing.org.uk/acd/dcs/overview.htm} and\\
\url{http://www.chilton-computing.org.uk/inf/literature/reports/alvey_report/overview.htm}} 

The UK {Alvey Programme} ran from 1983--87 and also invested in formal methods.%
\footnote{See for example Alvey News SE2/18 that contains a list of some of the projects funded by the Software Engineering Programme.}
The focus of the Alvey programme~\cite{oakley1990alvey} was pre-competitive advanced information technology research. 
 It comprised four areas that seemed particularly relevant at the time: 
 \begin{itemize}
 \item Software Engineering (led by David Talbot from ICL with Rob Witty from Informatics as his Deputy, who brought the focus on formal methods from DCS). 
 \item Intelligent Knowledge Based Systems 
  \item Man Machine Interaction 
  \item Advanced Microelectronics (VLSI Design) 
  \end{itemize}
  Research was a collaboration between academia, government and industry; it was directed into important areas and coordinated and the funding was substantial, GBP 350M at 1982 prices.  The Programme put together 210 projects lasting, on average, three years and involving 2500 people at its peak.
  
The largest Alvey project in {\em Software Engineering} funded Manchester University and ICL%
\footnote{This project came close to non-submission when,
having crafted a neat collaboration of three industrial organisations (STL, IDEC and ICL),
Jones  was informed that they were merging and that
the combination of their individual intended commitments was not defensible to a single board of directors.
Brian Warboys then of ICL steered a tense period of revisions at the eleventh hour.}
to construct an integrated project support environment dubbed {\em IPSE~2.5}~\cite{Snowdon90}.
Researchers at Manchester and EPSRC's own Rutherford Lab delivered a theorem proving assistant {\em Mural}~\cite{JJLM91}
which was licensed to {\em Winfrith Atomic Energy Establishment}.

Another activity under the Alvey programme led to the creation of a handbook of formal methods.
The contents attempted to identify areas of applicability for notations such as VDM, Z, CSP and CCS.


As recently as the 2010s,
formal methods was still identified as an area for growth of EPSRC funding.

Funding from the various European Union research framework programmes has also been a significant aid to formal methods research and deployment.
Again focussing on items where the authors have first-hand knowledge,
one of the longest-lasting impacts started with the funding of an activity called {\em VDM-Europe}:
meetings of experts in Brussels were supported for several years and led to the first symposium of  {\em VDM-Europe}~\cite{VDM87}.
These conferences morphed%
\footnote{Jim Woodcock tackled Jones about widening the remit of VDM-Europe to include other specification languages.}
into {\em FM-E}%
\footnote{see http://www.fmeurope.org/}
which, not only organises a highly-rated symposium at roughly 18-monthly intervals, 
but has also held two World Congresses~\cite{FM99,FMPorto}
and has widened its venues to North America, Singapore 
(and the 2021 event is planned for China).


\subsection{Tool support of formalism}


The question of how much the adoption of formal methods is influenced by the availability of tool support is interesting 
but is not universally agreed.
Early on,
large formal descriptions were constructed with minimal tool support.
A significant example is the formal description of PL/I from the IBM Vienna Lab.
It is probably fair to say that the risks of introducing inconsistencies are far higher when a document is revised
than when it is first constructed.
It is however clearly short-sighted not to at least syntax and type check any large block of formulae.

The question is how much further one can go without the tool support becoming an end in itself 
and possibly even distracting from the thought process that is crucial to the construction of an abstract model.
Jim Horning (private communication) captured one of the reservations about tools with his phrase
``mental versus metal tools''.
At least some of the differences in people's evaluation of the role of tools
can be accounted for by the contribution they hope to result from using formalism.

It is relatively easy to persuade organisations to use tools that analyse finished code in order to detect potential errors.
There are sub-issues here:
Peter O'Hearn 
(see Section~\ref{S-recent})
points out the tools that detect too many false positives are unlikely to endear themselves to developers.
But, broadly,
using model-checking tools in a development process that does not require deep understanding of formal methods from the developers
is an easier sell than starting out by insisting  that developers must employ formalism in the specification and early design phases.

The authors of the current paper, however, both argue that the real payoff of formal methods comes from their use early in the design phase.
Thomas has written~\cite{thomas1993industrial} about the cost-effectiveness of using formal methods early;
the {\em Tokeneer} study mentioned in Section~\ref{S-inside} supports this view;
Figure~\ref{F-CICS} provides evidence that formalism used to front-load thinking pays off;
a similar result can be seen in the dual-track study reported in~\cite{brookes1996formal}.

Many companies are willing to buy tools and see them as a quick fix --- 
but fail to recognise that tools exist to support methods and the main investment has to be in adopting the methods. 

The view of ``mental tools'' in no way removes the need for tool support
but it does moderate the extent to which the tool should be allowed to become the master of the method.
For example,
a large formal text might be regarded as an obvious input to a theorem proving system.
Unfortunately,
there are few success stories of such efforts.%
\footnote{The US work starting with the Boyer-Moore prover through to ACL/2 is a notable achievement
--- see \cite{Moore2019}.}
The reason would appear to be that theorem proving systems not only require learning another formalism but that their 
modes of interaction distract from thinking about the application in hand.
There are numerous stories of formal machine-checked proofs that do not actually capture what the user intended to establish.

Examples of tools that offer fairly direct support for established specification languages include
a VDM tool from Adelard%
\footnote{The work of the Adelard company would justify a paper of its own.
for example their development of the ``Dust Expert'' software is reported in~\cite{clement1999development}.}
and the IFAD Toolset (also for VDM) that has subsequently been developed into an open-source Overture tool.
Probably the most significant set of tools comes from Abrial, 
with the Rodin tool support%
\footnote{See http://www.event-b.org/}
for Event-B being the most recent.

\subsection{Standards}


One way in which the use of formalism could be encouraged is via standards.
%
In May 1989 the UK Ministry of Defence (MoD) issued a draft Interim Defence Standard 00-55 `Requirements for the procurement of safety critical software in defence equipment' for comment. The draft standard required that safety critical software should be formally specified and formally verified. Several companies in the defence industry attempted to persuade the MoD to withdraw the draft standard; but the MoD issued it as an interim standard in 1991 and made the standard mandatory for the SHOLIS project~\cite{king2000proof}. 
The interim standard was replaced by a full version in 1997,%
\footnote{\url{http://www.software-supportability.org/Docs/00-55_Part_1.pdf}} 
retaining the requirement for formal methods and including several examples from the SHOLIS project.  Def Stan 00-55 issue 3 recommended the use of civil standards such as RTSA DO-178, ISO 61508 and RTSA DO-254. 

IEC 61508 is the international standard for functional safety of programmable electronic systems. It requires that each safety function has a Safety Integrity Level that defines the allowable probability of failure: for continuous control of the most safety critical function (SIL4) the allowable probability must not exceed $10^{-8}$/hour.  The standard recommends  the use of formal methods for a SIL4 software safety function but does not mandate their use. In successive revisions of the standard, major European companies have repeatedly frustrated attempts to make formal methods mandatory for SIL 4 software.

Returning to Harlan Mills and what he achieved in IBM's Federal Systems Division
(and again relying on Jones' memory of personal discussions with Mills)
perhaps he had the best approach to standards.
At one point in time,
an FSD standard for software developers stated that programmers should accompany loops with
an indication of why they were claimed to achieve their aim;
there was not a mandated style for such annotations but the document did offer an example of a style that would serve.

There is of course also the question of standards for the formalism itself.
It was mentioned in Section~\ref{S-cons} that, 
during the CICS effort,
there was a request from IBM for a standard for the Z notation itself.
This is an understandable wish in that it opens up the possibility of sourcing tools and expertise from different organisations.
In fact in 1996, VDM was the first formal method notation to get an ISO standard%
\footnote{See https://www.iso.org/standard/22988.html}
and the Z standard followed in 2002.%
\footnote{See https://www.iso.org/standard/21573.html}




\section{Conclusions}
\label{S-concs}

Our emphasis in this paper has been on lessons that can be derived from early attempts 
to apply research on formal methods in significant software development.
This should in no way be seen as expressing reservations about the potential of the ideas
and Section~\ref{S-recent} points to two important recent success stories.
If we cannot learn from earlier difficulties,
no progress is made.
Rather than re-list all of the lessons noted earlier in the paper,
Section~\ref{S-final} pinpoints a few key messages.
It should also be mentioned that this paper was written on the expectation that it would be accompanied
by ones that relate to formalism in other European countries.

\subsection{More recent work}
\label{S-recent}

Recent attempts to use formal methods in industry include those spearheaded by 
Peter O'Hearn at Facebook and 
Byron Cook at {\em Amazon Web Services} (AWS).
O'Hearn made major contributions to {\em Concurrent Separation Logic}
(e.g.~\cite{OHearn07})
and went on to form, with colleagues, {\em Monoidics} which was acquired by Facebook in 2013.
The group has worked inside Facebook and~\cite{OHe15a} reports considerable success in creating tools
(see~\cite{DistefanoEt-19}) that are used 
in the standard development cycle by Facebook engineers.
In a private conversation with Jones, 
O'Hearn attributed the positive adoption by practicing engineers both to the creation of apposite tools and the fact that the
the general knowledge of fundamental computer science ideas is much more widespread now than in the attempts reported on earlier in this paper that mainly date from the last century.

Byron Cook's~\cite{Cook-18} is one of a sequence of papers reporting on application of formal methods at AWS;
his recorded keynote%
\footnote{https://www.youtube.com/watch?v=JfjLKBO27nw} 
talk at FLoC-18 in Oxford is inspirational
and, related to the lessons about management commitment,
even more telling are the talks from senior managers at AWS.%
\footnote{\url{https://www.youtube.com/watch?v=x6wsTFnU3eY}\\
\url{https://www.youtube.com/watch?v=BbXK_-b3DTk}}

\subsection{Lessons}
\label{S-final}

%
The contributions of UK researchers to the fundamental ideas that have shown how formal 
concepts and notations can be used in the description and development of software 
are significant.
Rather than list and attribute the scientific source material,
we have in this paper identified some significant attempts to deploy the theory into practical environments.
As indicated at the beginning of the paper,
we have mainly reported on deployments of which we have first hand knowledge.
These close encounters have made it possible to analyse the difficulties that were experienced.

Probably the most important single difficulty that complicated early deployments was the relatively 
small number of people available in receiving organisations who had acquaintance with a broad knowledge 
of theoretical concepts.
A short course on one or another specific notation does not fully equip someone to apply that notation to the key aim of abstracting away from 
the details of potential implementations;
similarly, nor does being shown a few examples of proofs convey the fundamental idea of recording a convincing correctness argument.
The more recent experiences from major companies like Microsoft, Amazon and Facebook suggest that 
the general educational environment now provides a far better basis than was available last century.

The attitude and commitment of management is clearly related and also of major importance.
The graph in Figure~\ref{F-CICS} indicates a challenge for managers who only feel comfortable when they can 
``weigh the code'':
just as in all engineering endeavours,
care and thought early in development clearly pays off later but does not yield  immediate lines of code.
(Nor does the drawing of careful architectural plans lay any bricks.)

Linking the points about technical expertise and management commitment is the issue of whether the key expertise is inside
the receiving organisation or supplied by external consultants.
Key advantages that come from the research expertise being within the deployment organisation are
bandwidth of communication and stability.
Perhaps more important is the avoidance of a split between ``practical'' work and {\em post hoc} formalisation.
This division appears to have been a source of problems in many of the early attempts to gain benefit from using formal methods in idustry.

Another issue is the choice of project --- perhaps crucially the first project.
A significant success story that is located about as far from our stated geographic focus as can be
is the work in Australia at {\em CSIRO~Data61} (previously known as NICTA).
Gerwin Klein and his team recognised the importance of microkernels because weaknesses here open 
any software built on top of them to subversion.
A recent paper on {\em sel4} is~\cite{HeiserEt-20} and earlier publications can be traced from its references.
In today's world where almost everything depends on software,
it is sometimes a financial aspect that identifies a development as ``business critical''.
There is, of course,
the class of ``safety critical'' systems that comprised early deployments of formal methods.
What the sel4 exercise is a reminder of is that underlying software can provide a Trojan Horse
on which reliance should only be proportional to its demonstrated trustworthiness.

One closing lesson
(and perhaps an uncomfortable one for the current authors) 
was mentioned by Jonathan Lawrence when he kindly reviewed the material relating to the IBM CICS project:
he suggested that one should ``not try to be too ambitious''.
It would be legitimate to ask whether some of the early deployments 
suffered precisely because researchers wanted to see the full extent of their research put into practice even if the 
receiving organisation was unready.

\section*{Acknowledgements}

The authors are indebted for input on aspects of this paper from:

\begin{itemize}

\item IBM/CICS: Tim Clement, Ian Hayes,  Steve King, Jonathan Lawrence and Carroll Morgan

\item STL: Tim Denvir, Mel Jackson

\item Praxis and Altran UK: David Bean, Roderick Chapman, Anthony Hall and Martyn Ould

\item UK Research Councils: Mel Jackson, David Duce and Rob Witty

\end{itemize}

\noindent
Troy Astarte offered extremely valuable comments on a draft of the paper.
The authors are grateful for the useful comments from the anonymous reviewers.
All remaining errors and opinions are however the responsibility of the authors.

Jones gratefully acknowledges the support for his research of grant 
RPG-2019-020
from the Leverhulme Foundation
and the EPSRC Strata Platform Grant.

\bibliography{master,C-extras}

\newcommand{\etalchar}[1]{$^{#1}$}
\providecommand{\noopsort}[1]{}
\begin{thebibliography}{DHJW85}

\bibitem[Abr96]{Abrial96}
J.-R. Abrial.
\newblock {\em {T}he {B}-{B}ook: {A}ssigning Programs to Meanings}.
\newblock Cambridge University Press, 1996.

\bibitem[Abr10a]{Abrial2010}
J.-R. Abrial.
\newblock {\em {M}odeling in {E}vent-{B}: {S}ystem and {S}oftware
  {E}ngineering}.
\newblock Cambridge University Press, New York, NY, USA, 2010.

\bibitem[Abr10b]{Abrial10}
J.-R. Abrial.
\newblock {\em {T}he {E}vent-{B} {B}ook}.
\newblock Cambridge University Press, Cambridge, UK, 2010.

\bibitem[Ame02]{amey2002correctness}
Peter Amey.
\newblock Correctness by construction: Better can also be cheaper.
\newblock {\em CrossTalk: the Journal of Defense Software Engineering},
  2:24--28, 2002.

\bibitem[AO19]{AptOlderog-19}
Krzysztof~R. Apt and Ernst-R{\"u}diger Olderog.
\newblock Fifty years of {Hoare's} logic.
\newblock {\em Formal Aspects of Computing}, 31(6):751--807, 2019.

\bibitem[Ast19]{AstartePhD}
Troy~K. Astarte.
\newblock {\em Formalising Meaning: a History of Programming Language
  Semantics}.
\newblock PhD thesis, Newcastle University, 6 2019.

\bibitem[BBG{\etalchar{+}}60]{Naur60}
John~W. Backus, Friedrich~L. Bauer, Julien Green, Charles Katz, John McCarthy,
  Peter Naur, Alan~J. Perlis, Heinz Rutishauser, Klaus Samelson, Bernard
  Vauquois, et~al.
\newblock Report on the algorithmic language {ALGOL 60}.
\newblock {\em Numerische Mathematik}, 2(1):106--136, 1960.

\bibitem[BFL96]{brookes1996formal}
TM~Brookes, John~S Fitzgerald, and Peter~Gorm Larsen.
\newblock Formal and informal specifications of a secure system component:
  final results in a comparative study.
\newblock In {\em International Symposium of Formal Methods Europe}, pages
  214--227. Springer-Verlag, 1996.

\bibitem[BGJ06]{DIRC-Structure06}
D.~Besnard, C.~Gacek, and C.~B. Jones, editors.
\newblock {\em Structure for Dependability: Computer-Based Systems from an
  Interdisciplinary Perspective}.
\newblock Springer-Verlag, 2006.

\bibitem[BHP83]{beichter1983slan}
F~Beichter, Otthein Herzog, and Heiko Petzsch.
\newblock Slan-4: a language for the specification and design of large software
  systems.
\newblock {\em IBM journal of research and development}, 27(6):558--576, 1983.

\bibitem[BJ78]{BjornerJones78}
D.~Bj{\o}rner and C.~B. Jones, editors.
\newblock {\em The Vienna Development Method: The Meta-Language}, volume~61 of
  {\em Lecture Notes in Computer Science}.
\newblock Springer-Verlag, 1978.

\bibitem[BJ82]{BjornerJones82}
Dines Bj{\o}rner and Cliff~B. Jones, editors.
\newblock {\em Formal Specification and Software Development}.
\newblock Prentice Hall International, 1982.

\bibitem[BJMN87]{VDM87}
Dines Bj{\o}rner, C.~B. Jones, M.~{Mac an Airchinnigh}, and E.~J. Neuhold,
  editors.
\newblock {\em VDM -- A Formal Definition at Work}, volume 252 of {\em Lecture
  Notes in Computer Science}. Springer-Verlag, 1987.

\bibitem[BJW06]{Barnes-06b}
J.~Barnes, R.~Johnson, and J.~C. Widmaier.
\newblock Engineering the {Tokeneer} enclave protection software.
\newblock {\em Proceedings of IEEE International Symposium on Secure Software
  Engineering}, 2006.

\bibitem[BvW98]{BackWright98}
R.-J.~R. Back and J.~von Wright.
\newblock {\em Refinement Calculus: A Systematic Introduction}.
\newblock Springer-Verlag, New York, 1998.

\bibitem[CCFJ99]{clement1999development}
Tim Clement, Ian Cottam, Peter Froome, and Claire Jones.
\newblock The development of a commercial ``shrink-wrapped application'' to
  safety integrity level 2: The dust-expert{\texttrademark} story.
\newblock In {\em International Conference on Computer Safety, Reliability, and
  Security}, pages 216--225. Springer-Verlag, 1999.

\bibitem[CK85]{Cam85}
Martin Campbell-Kelly.
\newblock Christopher {Strachey}, 1916-1975: A biographical note.
\newblock {\em IEEE Annals of the History of Computing}, 1(7):19--42, 1985.

\bibitem[CNS87]{TR12.260}
B.P. Collins, J.E. Nicholls, and I.H. Sorensen.
\newblock Introducing formal methods: The {CICS} experience with {Z}.
\newblock Technical Report TR12.060, IBM Hursley Laboratory, 12 1987.

\bibitem[Coh88]{Cohn88}
A.~Cohn.
\newblock A proof of correctness of the {VIPER} microprocessor: The first
  level.
\newblock In G.~Birtwistle and P.A. Subrahmanyam, editors, {\em VLSI
  Specification, Verification and Synthesis}, volume~35. Kluwer, 1988.

\bibitem[Coo18]{Cook-18}
Byron Cook.
\newblock Formal reasoning about the security of {Amazon Web Services}.
\newblock In {\em International Conference on Computer Aided Verification},
  pages 38--47. Springer-Verlag, 2018.

\bibitem[CS14]{ChapmanSchanda-14}
R.~Chapman and F.~Schanda.
\newblock Are we there yet? 20 years of industrial theorem proving with
  {SPARK}.
\newblock In {\em Proceedings of Interactive Theorem Proving (ITP)}, volume
  8558 of {\em Lecture Notes in Computer Science}, pages 17--26.
  Springer-Verlag, 2014.

\bibitem[Den17]{Denvir-17}
T.~Denvir.
\newblock Fifty years of formal methods in software engineering.
\newblock {\em FACTS}, 2017.

\bibitem[DFLO19]{DistefanoEt-19}
Dino Distefano, Manuel F\"{a}hndrich, Francesco Logozzo, and Peter~W. O'Hearn.
\newblock Scaling static analyses at {Facebook}.
\newblock {\em Communications of the ACM}, 62(8):62--70, 2019.

\bibitem[DHJW85]{Denvir85}
B.~T. Denvir, W.~T. Harwood, M.~I. Jackson, and M.~J. Wray.
\newblock {\em The Analysis of Concurrent Systems: Cambridge, September 1983,
  Proceedings of a Workshop}, volume 207 of {\em Lecture Notes in Computer
  Science}.
\newblock Springer Verlag, Berlin, 1985.

\bibitem[DM94]{B4RATP}
Babak Dehbonei and Fernando Mejia.
\newblock Formal methods in the railways signalling industry.
\newblock In {\em International Symposium of Formal Methods Europe}, volume 873
  of {\em Lecture Notes in Computer Science}, pages 26--34. Springer-Verlag,
  1994.

\bibitem[Fen95]{Fenney-95}
T~Fenney.
\newblock {Sizewell B Primary Protection System}. an assessment of the
  confirmatory review processes.
\newblock Technical report, Nuclear Electric plc, 1995.

\bibitem[GMW79]{Gordon79}
M.~Gordon, R.~Milner, and C.~Wadsworth.
\newblock {\em Edinburgh LCF}, volume~78 of {\em Lecture Notes in Computer
  Science}.
\newblock Springer-Verlag, 1979.

\bibitem[Gor00]{gordon2000lcf}
Mike Gordon.
\newblock From {LCF} to {HOL}: a short history.
\newblock In {\em Proof, language, and interaction}, pages 169--186, 2000.

\bibitem[Hal96]{Hall96}
Anthony Hall.
\newblock Using formal methods to develop an {ATC} information system.
\newblock {\em IEEE Software}, 13(2):66--76, 1996.
\newblock Hard copy.

\bibitem[Hal05]{Hall-05}
Anthony Hall.
\newblock Realising the benefits of formal methods.
\newblock In {\em Formal Methods and Software Engineering}, volume 3785 of {\em
  Lecture Notes in Computer Science}, pages 1--4. Springer-Verlag, 2005.

\bibitem[Hay87]{Hayes87a}
I.~J. Hayes, editor.
\newblock {\em Specification Case Studies}.
\newblock Prentice Hall International, 1987.

\bibitem[Hay92]{Hayes92a}
I.~J. Hayes, editor.
\newblock {\em Specification Case Studies}.
\newblock Prentice Hall International, second edition, 1992.

\bibitem[Hay93]{Hayes93}
Ian Hayes, editor.
\newblock {\em Specification Case Studies}.
\newblock Prentice Hall International, Englewood Cliffs, N.J., USA, second
  edition, 1993.

\bibitem[HC02a]{HallChapman-02}
A.~Hall and R.~Chapman.
\newblock Correctness by construction: building a commercial secure system.
\newblock {\em Software}, 19(1):18--25, 2002.

\bibitem[HC02b]{hall2002correctness}
Anthony Hall and Roderick Chapman.
\newblock Correctness by construction: Developing a commercial secure system.
\newblock {\em IEEE software}, 19(1):18--25, 2002.

\bibitem[HDNS96]{hoareJ1996applying}
Jonathan Hoare, Jeremy Dick, Dave Neilson, and Ib~S{\o}rensen.
\newblock Applying the {B} technologies to {CICS}.
\newblock In {\em International Symposium of Formal Methods Europe}, pages
  74--84. Springer-Verlag, 1996.

\bibitem[HK91]{HoustonKing}
Iain Houston and Steve King.
\newblock {CICS} project report experiences and results from the use of z in
  ibm.
\newblock In {\em International Symposium of VDM Europe}, pages 588--596.
  Springer-Verlag, 1991.

\bibitem[HKA20]{HeiserEt-20}
Gernot Heiser, Gerwin Klein, and June Andronick.
\newblock sel4 in australia: from research to real-world trustworthy systems.
\newblock {\em Communications of the ACM}, 63(4):72--75, 2020.

\bibitem[Hoa69]{Hoare69a}
C.~A.~R. Hoare.
\newblock An axiomatic basis for computer programming.
\newblock {\em Communications of the ACM}, 12(10):576--580, 1969.

\bibitem[Hoa85]{Hoare85d}
C.~A.~R. Hoare.
\newblock {\em Communicating Sequential Processes}.
\newblock Prentice Hall, 1985.

\bibitem[JA16]{JonesAstarte-16TR}
Cliff~B. Jones and Troy~K. Astarte.
\newblock {An exegesis of four formal descriptions of {ALGOL} 60}.
\newblock Technical Report CS-TR-1498, Newcastle University School of Computer
  Science, 9 2016.

\bibitem[JA18]{JonesAstarte-SC-proc}
Cliff~B. Jones and Troy~K. Astarte.
\newblock Challenges for semantic description: comparing responses from the
  main approaches.
\newblock In Jonathan~P. Bowen, Zili Zhang, and Zhiming Liu, editors, {\em
  Proceedings of the Third School on Engineering Trustworthy Software Systems},
  volume 11174 of {\em Lecture Notes in Computer Science}, pages 176--217,
  2018.

\bibitem[Jac75]{Jackson75}
Michael Jackson.
\newblock {\em Principles of Program Design}.
\newblock Academic Press, 1975.

\bibitem[Jac00]{Jackson00}
Michael Jackson.
\newblock {\em Problem Frames: Analyzing and Structuring Software Development
  Problems}.
\newblock Addison-Wesley, 2000.

\bibitem[JDS85]{JacksonDenvirShaw-85}
M.~I. Jackson, B.~T. Denvir, and R.~C. Shaw.
\newblock Experience of introducing the vienna development method into an
  industrial organisation.
\newblock In {\em Proceedings of the TAPSOFT Conference}, number 186 in lncs.
  Springer, 1985.

\bibitem[JJLM91]{JJLM91}
C.~B. Jones, K.~D. Jones, P.~A. Lindsay, and R.~Moore.
\newblock {\em {mural}: A Formal Development Support System}.
\newblock Springer-Verlag, 1991.

\bibitem[Jon71]{Jones71c}
C.~B. Jones.
\newblock Development of correct programs: An example based on {Earley's}
  recogniser.
\newblock Technical Report TN 9000, IBM Laboratory, Hursley, 4 1971.

\bibitem[Jon73]{Jones73a}
C.~B. Jones.
\newblock Formal development of programs.
\newblock Technical Report 12.117, IBM Laboratory Hursley, 6 1973.

\bibitem[Jon80]{Jones80a}
C.~B. Jones.
\newblock {\em Software Development: A Rigorous Approach}.
\newblock Prentice Hall International, Englewood Cliffs, N.J., USA, 1980.

\bibitem[Jon01]{Jones01d}
C.~B. Jones.
\newblock The transition from {VDL} to {VDM}.
\newblock {\em Journal of Universal Computer Science}, 7(8):631--640, 2001.

\bibitem[Jon03]{Jones03i}
Cliff~B. Jones.
\newblock The early search for tractable ways of reasoning about programs.
\newblock {\em IEEE Annals of the History of Computing}, 25(2):26--49, 2003.

\bibitem[KHCP00]{king2000proof}
Steve King, Jonathan Hammond, Rod Chapman, and Andy Pryor.
\newblock Is proof more cost-effective than testing?
\newblock {\em IEEE Transactions on software Engineering}, 26(8):675--686,
  2000.

\bibitem[LW69]{LucasWalk69}
Peter Lucas and Kurt Walk.
\newblock On the formal description of {PL/I}.
\newblock {\em Annual Review in Automatic Programming}, 6:105--182, 1969.

\bibitem[Mac01]{Mac01}
Donald MacKenzie.
\newblock {\em Mechanizing Proof: Computing, Risk, and Trust}.
\newblock MIT Press, 2001.

\bibitem[MC93]{Ella-93}
J.D. Morison and A.S. Clarke.
\newblock {\em Ella 2000: A Language for Electronic System Design}.
\newblock McGraw Hill, 1993.

\bibitem[McC66]{McC66}
John McCarthy.
\newblock A formal description of a subset of {ALGOL}.
\newblock In {\em Formal Language Description Languages for Computer
  Programming}, pages 1--12. North-Holland, 1966.

\bibitem[Mid89]{middelburg1989vvsl}
Cornelis~Adam Middelburg.
\newblock Vvsl: A language for structured vdm specifications.
\newblock {\em Formal aspects of computing}, 1(1):115--135, 1989.

\bibitem[Mid90]{Middelburg90}
C.~A. Middelburg.
\newblock {\em Syntax and Semantics of VVSL: A Language for Structured VDM
  Specifications}.
\newblock PhD thesis, PTT Research, Leidschendam, Department of Applied
  Computer Science, 9 1990.

\bibitem[Mil80]{Milner80a}
R.~Milner.
\newblock {\em A Calculus of Communicating Systems}, volume~92 of {\em Lecture
  Notes in Computer Science}.
\newblock Springer-Verlag, 1980.

\bibitem[MO94]{MarshONeil1994}
Marsh and O'Neil.
\newblock The formal semantics of {SPARK}, 1994.

\bibitem[Moo19]{Moore2019}
J~Strother Moore.
\newblock Milestones from the {Pure Lisp} theorem prover to {ACL2}.
\newblock {\em Formal Aspects of Computing}, 31(6):699--732, 2019.

\bibitem[Mor94]{Morgan94}
C.~C. Morgan.
\newblock {\em Programming from Specifications}.
\newblock Prentice Hall, second edition, 1994.

\bibitem[MP93]{mcmorran1993}
Mike McMorran and Steve Powell.
\newblock {\em Z Guide for Beginners}.
\newblock Alfred Waller Limited, 1993.

\bibitem[O'H07]{OHearn07}
P.~W. O'Hearn.
\newblock Resources, concurrency and local reasoning.
\newblock {\em Theoretical Computer Science}, 375(1-3):271--307, 5 2007.

\bibitem[O'H15]{OHe15a}
Peter O'Hearn.
\newblock From categorical logic to {Facebook} engineering.
\newblock In {\em Logic in Computer Science (LICS), 2015 30th Annual ACM/IEEE
  Symposium on}, pages 17--20. IEEE, 2015.

\bibitem[OO90]{oakley1990alvey}
Brian Oakley and Kenneth Owen.
\newblock {\em Alvey: Britain's strategic computing initiative}.
\newblock MIT press, 1990.

\bibitem[Pau18]{Gordon-obit}
Lawrence~C Paulson.
\newblock {Michael John Caldwell Gordon (FRS 1994)}, 28 {February} 1948--22
  {August} 2017.
\newblock {\em arXiv preprint arXiv:1806.04002}, 2018.
\newblock Royal Society obituary.

\bibitem[Pei91]{peirce1991peirce}
Charles~Sanders Peirce.
\newblock {\em Peirce on signs: Writings on semiotic}.
\newblock UNC Press Books, 1991.

\bibitem[PH97]{PfleegerEt-97}
S.L. Pfleeger and L.~Hatton.
\newblock Investigating the influence of formal methods.
\newblock {\em Computer}, pages 33--43, 1997.

\bibitem[Plo81]{Plotkin81}
G.~D. Plotkin.
\newblock A structural approach to operational semantics.
\newblock Technical Report DAIMI FN-19, Aarhus University, 1981.

\bibitem[Plo04a]{Plotkin03b}
Gordon~D. Plotkin.
\newblock The origins of structural operational semantics.
\newblock {\em Journal of Logic and Algebraic Programming}, 60--61:3--15,
  July--December 2004.

\bibitem[Plo04b]{Plotkin03a}
Gordon~D. Plotkin.
\newblock A structural approach to operational semantics.
\newblock {\em Journal of Logic and Algebraic Programming}, 60--61:17--139,
  July--December 2004.

\bibitem[PNW19]{Paulson2019}
Lawrence~C. Paulson, Tobias Nipkow, and Makarius Wenzel.
\newblock From {LCF} to {Isabelle/HOL}.
\newblock {\em Formal Aspects of Computing}, 31(6):675--698, 2019.

\bibitem[Ros77]{SADT}
D.T. Ross.
\newblock Structured analysis (sa): A language for communicating ideas.
\newblock {\em IEEE Transactions on Software Engineering}, SE-3(1):16--34,
  1977.

\bibitem[RT13]{romanovsky2013industrial}
Alexander Romanovsky and Martyn Thomas.
\newblock {\em Industrial deployment of system engineering methods}.
\newblock Springer-Verlag, 2013.

\bibitem[Sno90]{Snowdon90}
R.A. Snowdon.
\newblock An introduction to the ipse 2.5 project.
\newblock In F.~Long, editor, {\em Software Engineering Environments}, volume
  467 of {\em lncs}. springer, 1990.

\bibitem[Spi88]{Spivey88}
J.~M. Spivey.
\newblock {\em Understanding Z---A Specification Language and its Formal
  Semantics}.
\newblock Cambridge Tracts in Computer Science 3. Cambridge University Press,
  1988.

\bibitem[Spi89]{Spivey89}
J.~M. Spivey.
\newblock {\em The Z Notation: A Reference Manual}.
\newblock Prentice Hall International, 1989.

\bibitem[Spi92]{Spivey92}
J.M. Spivey.
\newblock {\em The {Z} Notation: A Reference Manual}.
\newblock Prentice Hall International, second edition, 1992.

\bibitem[Ste66]{Steel66}
T.~B. Steel, editor.
\newblock {\em Formal Language Description Languages for Computer Programming}.
\newblock North-Holland, 1966.

\bibitem[Sys06]{Spark-Hercules}
Praxis~Critical Systems.
\newblock Spark: A successful contribution to the {Lockheed C130-J Hercules
  II}.
\newblock Praxis report, 12 2006.

\bibitem[tBMO19]{FMPorto}
Maurice~H. ter Beek, Annabelle McIver, and Jos{\'{e}}~N. Oliveira, editors.
\newblock {\em Formal Methods - The Next 30 Years - Third World Congress, {FM}
  2019, Porto, Portugal, October 7-11, 2019, Proceedings}, volume 11800 of {\em
  Lecture Notes in Computer Science}. Springer-Verlag, 2019.

\bibitem[Tho93]{thomas1993industrial}
Martyn Thomas.
\newblock The industrial use of formal methods.
\newblock {\em Microprocessors and Microsystems}, 17(1):31--36, 1993.

\bibitem[War93]{Ward-93}
NJ~Ward.
\newblock The rigorous retrospective static analysis of the {Sizewell `B'
  Primary Protection System} software.
\newblock In {\em SAFECOMP'93}, pages 171--181. Springer-Verlag, 1993.

\bibitem[WB83]{WoodwardBond-83}
P.M. Woodward and S.G. Bond.
\newblock {\em Guide to Algol 68 for users of RS Systems}.
\newblock Edward Arnold, 1983.

\bibitem[WB92]{woodcock1992w}
J.C.P. Woodcock and S.M. Brien.
\newblock {W}: a logic for {Z}.
\newblock In {\em Z User Workshop, York 1991}, pages 77--96. Springer-Verlag,
  1992.
\newblock Hard copy.

\bibitem[WD96]{woodcock1996using}
Jim Woodcock and Jim Davies.
\newblock {\em Using {Z}: Specification, Refinement and Proof}.
\newblock Prentice Hall International, 1996.

\bibitem[WGC10]{WoodcockEt-10}
J.~Woodcock, E.~{G{\"o}kce Aydal}, and R.~Chapman.
\newblock The {Tokeneer} experiments.
\newblock In C.~B.~Jones et~al, editor, {\em Reflections on the Work on C. A.
  R. Hoare}, pages 405--430. Springer-Verlag, 2010.

\bibitem[Wir77]{Wirth77}
Niklaus Wirth.
\newblock What can we do about the unnecessary diversity of notation for
  syntactic definitions?
\newblock {\em Commun. ACM}, 20(11):822--823, 1977.

\bibitem[WLBF09]{WoodcockEt09}
J.~Woodcock, P.~G. Larsen, J.~Bicarregui, and J.~Fitzgerald.
\newblock Formal methods: Practice and experience.
\newblock {\em ACM Computing Surveys}, 41(4), 10 2009.

\bibitem[WMC17]{WhiteEt-17}
N.~White, S.~Matthews, and R.~Chapman.
\newblock Formal verification: will the seedling ever flower?
\newblock {\em Phil Trans R Soc A}, 375(2104), 2017.

\bibitem[Wor92]{Wordsworth92}
J.~B. Wordsworth.
\newblock {\em Software Development with {Z}}.
\newblock Addison-Wesley, 1992.

\bibitem[WWD99]{FM99}
J.M. Wing, J.~Woodcock, and J.~Davies, editors.
\newblock {\em FM'99 -- Formal Methods FM'99 -- Formal Methods}, volume 1709.
  Springer-Verlag, 1999.

\end{thebibliography}
	 
\end{document}